\documentclass[prl,amsmath,amssymb,twocolumn,showpacs]{revtex4}
\usepackage{graphicx}
\begin{document}
\title{Quantitative Test of $SO(5)$ Symmetry in the Vortex State of $Nd_{1.85}Ce_{0.15}CuO_4$}
\author{Han-Dong Chen}
\affiliation{Department of Applied Physics, McCullough building, Stanford University, Stanford, CA 94305}
\author{Congjun Wu}
\affiliation{Department of Physics, McCullough building, Stanford University, Stanford, CA 94305}
\author{Shou-Cheng Zhang}
\affiliation{Department of Physics, McCullough building, Stanford University, Stanford, CA 94305}

\begin{abstract}
By numerically solving models with competing superconducting and
antiferromagnetic orders, we study the magnetic field dependence
of the antiferromagnetic moment in both the weak and strong field
regimes. Through a comparison with the neutron scattering results of Kang
and Matsuura {\it et al}. on $Nd_{1.85}Ce_{0.15}CuO_4$, we
conclude that this system is close to a SO(5) symmetric critical
point. We also make a quantitative prediction on increasing the
upper critical field $B_{c2}$ and the superconducting transition
temperature $T_c$ by applying an in-plane magnetic field.
\end{abstract}
\pacs{74.20.-z, 74.20.De, 74.25.Op, 74.25.Ha} \maketitle 
A central question in the field of high $T_c$ superconductivity concerns the
nature of the competing state and its quantum phase transition to
the superconducting (SC) state\cite{levi2002,sachdev2002A}. The
$SO(5)$ theory\cite{zhang1997} considers the antiferromagnetic
(AF) state to be the dominant competing state, and describes its
competition with the SC state in terms of an unified theory with
approximate $SO(5)$ symmetry. This theory predicts that the
competing AF order reveals itself in the SC vortex
core\cite{zhang1997,arovas1997}. This prediction has now been
verified experimentally in a number of different compounds and by
a variety of different experimental
techniques\cite{katano2000,matsuura2003,Lake2002,Lake2001,Miller2001,Mitrovic2001,Khaykovich2001,Mitrovic2002,KUMAGAI2003A,Kang2003}.
Experimental and theoretical progress on the AF vortex lattice is
reviewed in recent articles\cite{levi2002, sachdev2002A}.

While the experimental observation of enhanced AF correlation in
the vortex state is encouraging, the limitation of low magnetic
field could not convincingly establish the AF state as the
competing order. Only recently, it became possible to probe the
magnetic structure of the AF vortex core with magnetic fields
comparable to the upper critical field $B_{c2}$. Kang {\it et
al}\cite{Kang2003} and Matsuura {\it et al}\cite{matsuura2003}
performed neutron scattering experiment on the electron-doped
$Nd_{2-x}Ce_xCuO_{4-\delta}$ (NCCO) crystal under a magnetic field
of $16T$, beyond the upper critical field $B_{c2}=6.2T$ of this
material. The experiment finds a field induced AF scattering at
the commensurate wave vector $(\pi,\pi,0)$, where the AF moment
scales approximately linearly up to $B_{c2}$. The AF moment
decreases with increasing field beyond $B_{c2}$. The experiment
also finds field induced scattering at the $(\pi,0,0)$ position.
Scattering at this position has a different field dependence at
high field, and may not be intrinsically related to the
$(\pi,\pi,0)$ scattering. We shall not discuss this scattering
position further in this paper.

The wealth of the experimental data on this system provides an
opportunity to quantitatively test theoretical models. In this
work, we use the Landau-Ginzburg model of competing AF and SC order,
including the Zeeman term. By quantitatively fitting the
parameters to the experimentally measured field dependence of the
AF moment, we could determine how close this system is to a
$SO(5)$ symmetric point. We show that the departure from the
$SO(5)$ symmetric parameters in general leads to departure from
the linear dependence of the AF moment on the magnetic field,
in the high field regime.
Therefore, the experimentally observed linear dependence up to
$B_{c2}$ directly implies that $Nd_{1.85}Ce_{0.15}CuO_4$ is close
to a $SO(5)$ symmetric point. The magnetic field has a dual effect
on the AF moment. On one hand, it creates AF moment in the vortex
core, on the other hand, it reduces the AF moment by canting the
spins uniformly towards to field direction. This explains the
reduction of the AF moment when the magnetic field exceeds
$B_{c2}$.  We also show that the SC coherence length is
self-consistently determined by the interaction of the SC and AF
order parameters, therefore, $B_{c2}$ is reduced in this system
from its bare value. From this observation we predict that the
value of $B_{c2}$ would be enhanced, if the AF order is suppressed
by an in-plane magnetic field. The same reasoning lead to an
enhancement of $T_c$ in a system where AF and SC coexist
uniformly.

A minimal GL free energy that describes the competition of AF and
SC orders under magnetic field takes the form
\begin{eqnarray}
  \mathcal{F}=\mathcal{F}_\psi+\mathcal{F}_m+\mathcal{F}_{int}
  + \kappa^2 b^2
  \label{F}
\end{eqnarray}
with
\begin{subequations}
\begin{eqnarray}
  \mathcal{F}_\psi&=&\rho_1\left|\left({\bf \nabla}-i~2\pi
  {\bf A}\right)\psi\right|^2
  + r_1|\psi|^2 +\frac{u_1}{2}|\psi|^4\nonumber\\ ~\\
  \mathcal{F}_m&=&\rho_2|{\bf \nabla m}|^2 + r_2|{\bf m}|^2
  +\frac{u_2}{2}|{\bf m}|^4\nonumber\\
  &&+\chi~ b^2|{\bf m}_\perp|^2 \\
  \mathcal{F}_{int}&=&u_{12}|\psi|^2|{\bf m}|^2 .
\end{eqnarray}
\end{subequations}
Here, we express the free energy and the order parameters in
dimensionless form. $\mathcal{F}_\psi$ is the standard GL free
energy of the SC order parameter $\psi$, $\mathcal{F}_m$ is the
free energy of the AF order parameter ${\bf m}$
\cite{Kosterlitz1976}. ${\bf A}$ is the vector potential and ${\bf
b}=({\bf \nabla}\times {\bf A})\times a^2$ is the dimensionless magnetic field
measured in units of $\phi_0/a^2$, where $\phi_0=hc/2e$ is the
London flux quantum and $a$ is the lattice constant. ${\bf
m}_\perp$ is the component of the AF moment perpendicular to the
magnetic field. $\kappa$ is the GL parameter. $\mu_B$ is the Bohr
magneton. The $\chi$ term describes the Zeeman coupling of the AF
moment with the external magnetic field. The interaction term
$u_{12}$ is the key term describing the interaction between
the AF and the SC order parameters. When
$u^2_{12}>u_2u_1$, there is a direct first
order transition between the AF and the SC state. On the other
hand, when $u^2_{12}<u_2u_1$, the system
undergoes two second order phase transitions, with an intermediate
phase where AF and SC coexist uniformly. In the special case of
$u^2_{12}=u_2u_1$, the potential can be
scaled into a $SO(5)$ symmetric form at the quantum critical point
where $r_2/\sqrt{u_2}=r_1/\sqrt{u_1}$, and
the AF and SC orders can be freely rotated into each other.
Therefore, the relationship among the three quartic coefficients
provides a crucial test on the $SO(5)$ symmetry of the model. On
the other hand, by tuning the doping and the chemical potential,
the quadratic coefficients can always be tuned to reach the
quantum critical points. By quantitatively fitting the neutron
scattering data in the vortex state of $Nd_{1.85}Ce_{0.15}CuO_4$,
we can determine the value
$u_{12}/\sqrt{u_2u_1}$, and consequently the
nature of the quantum phase transition between AF and SC.

We use the relaxation method\cite{{alder1984},{xu1996}} to find
the numerical solution that minimizes the GL free energy
(\ref{F}). We assume the system is an extreme type-II
superconductor, $\kappa\gg1$, so that the magnetic field is
uniformly distributed over the sample. We choose the symmetrical
gauge so that the vector potential ${\bf A}$ reads
\begin{eqnarray}
  {\bf A} = -\frac{{\bf r}\times {\bf b}}{2}.
\end{eqnarray}
The boundary conditions are then given by
\begin{subequations}
\begin{eqnarray}
  \psi({\bf r}+L_x\hat{{\bf x}})&=&\psi({\bf r})e^{-i 2\pi A_x L_x/a }
  \label{bound-condition-start} \\
  \psi({\bf r}+L_y\hat{{\bf y}})&=&\psi({\bf r})e^{-i 2\pi A_y L_y/a}\\
  m({\bf r}+L_x\hat{{\bf x}})&=&m({\bf r})\\
  m({\bf r}+L_y\hat{{\bf y}})&=&m({\bf r}).  \label{bound-condition-end}
\end{eqnarray}
\end{subequations}
Here, $L_x$ and $L_y$ are the width and height of the magnetic
unit cell that satisfy the condition $(b_z
\phi_0/a^2)L_xL_y=\phi_0$. $A_x$ and $A_y$ are the components of
the vector potential ${\bf A}$ in the $ab$-plane. With the free
energy (\ref{F}) and the boundary conditions
(\ref{bound-condition-start})-(\ref{bound-condition-end}), we
choose $\psi$, $\psi^*$ and $m$ as independent variables. The
relaxation iteration equations read
\begin{subequations}
\begin{eqnarray}
  \psi^{(n+1)}&=&\psi^{(n)}-\delta_1~\frac{\partial \mathcal{F}}{\partial \psi^*}
  \biggr|^{(n)}\\
  m^{(n+1)}&=&m^{(n)}-\delta_2~\frac{\partial \mathcal{F}}{\partial m}
  \biggr|^{(n)},
\end{eqnarray}
\end{subequations}
where $\delta_1$ and $\delta_2$ are small positive numbers
adjusted to optimize the convergence and $n$ denotes the
generation of the iteration. Starting from a proper initial state,
this relaxation procedure quickly reaches convergence. We then
repeat the relaxation procedure for different sizes of magnetic
unit cell, or equivalently for different magnetic fields, to
obtain the field dependence of the AF moment.

\begin{figure}[t]
    \includegraphics[scale=0.8]{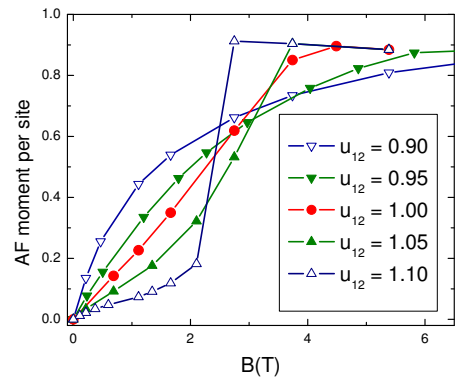}
    \caption{The plot of field dependence of AF moment for differen parameters.
    The other parameters are $\rho_1=\rho_2= a^2$, $r_1=-1$,
$r_2=-0.85$, $u_1=u_2=1$ and $\chi=42.4$. Here, the parameters are
chosen such that the maximum SC order is $1$ and the SC coherence
length at zero field equals the lattice constant $a$ of our
discrete model. }\label{Fig-AF-moment}
\end{figure}

When $u_{12}=0$, the two order parameters decouple.
The bare SC coherence length is given by
$\xi^2_0=\rho_1/|r_1|$, and the bare upper critical
field is given by $B_0=\phi_0/2\pi\xi^2_0$. However, when
$u_{12}>0$, the magnetic field induces AF moment in the
vortex core regions, which renormalizes the coherence length and
the upper critical field to $\xi>\xi_0$ and $B_{c2}<B_0$. These
values are self-consistently determined by solving the coupled GL
equations in the vortex state. Fig.\ref{Fig-AF-moment} shows the
field dependence of the AF moment, for different values of
$u_{12}$. We see that this value directly determines the
curvature of the field dependence curve. For strong repulsion,
$u^2_{12}>u_2u_1$, the AF moment increases
slowly at low field since the AF moment is strongly
suppressed by the SC order in the bulk. At high field close to
$B_{c2}$, the AF moment increases steeply since most of the SC
order are destroyed. Therefore, the field dependence of the AF
moment has a positive curvature in this
regime. In contrast, when $u^2_{12}<u_2u_1$,
the field dependence is reversed. The AF moment increases steeply
for low field, and levels off at high field, leading to a negative
curvature. In the $SO(5)$ symmetric case where
$u^2_{12}=u_2u_1$, the AF moment scales
linearly with $B$ up to $B_{c2}$.

The linear dependence was first predicted by Arovas et
al\cite{arovas1997} in the low field regime, based on the argument
that the AF moment scales with the number of vortices, which is
linearly proportional to the field. Here we have shown that this
linear dependence generally does not hold for fields close to
$B_{c2}$, unless the system is $SO(5)$ symmetric. Demler {\it et al.}
\cite{demler2001} showed that the circulating currents around the
vortices lead to a logarithmic correction of the form $B\ln(B_{c2}/B)$.
Their result is derived in the weak field regime
$B\ll B_{c2}$. This behavior leads to an infinite slope at $B=0$.
The experimentally measured field dependence shown in Fig. 3e of
reference \cite{Kang2003} does not have this feature. Zhang {\it et 
al.}\cite{ZHANG2002} studied the similar problem, in the $B<B_{c2}$ regime, 
mostly for the parameters where $u^2_{12}<u_1u_2$, and found the field dependence 
of the induced magnetic moment to have a negative curvature. This result is  
consistent with ours in the same parameter regime.

\begin{figure}[t]
    \includegraphics[scale=0.85]{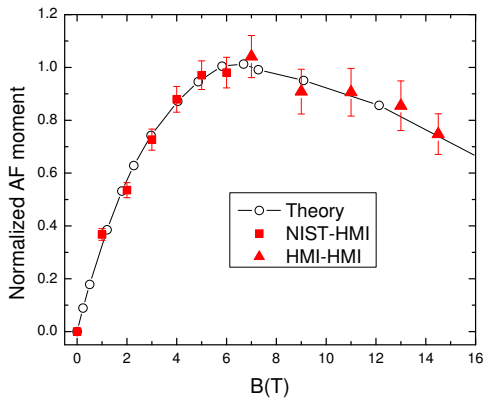}
    \caption{The field dependence of the AF moment. 
    The solid squares are obtained by subtracting the intensity of $(0,0,4.4\pi)$
    measured at Hahn-Meitner-Institute (HMI)\cite{matsuura-unpublished} from
    the intensity of $(\pi,\pi,0)$ measured at
    National Institute of Standards and Technology\cite{matsuura2003}. 
    The solid triangles are obtained subtracting the intensity of $(0,0,4.4\pi)$
    measured at HMI\cite{matsuura-unpublished} from the intensity of $(\pi,\pi,0)$
    measured at HMI\cite{matsuura2003}. The parameters used in obtaining
    theoretical curve are $\rho_1=\rho_2= a^2$, $r_1=-1$, $r_2=-0.85$,
    $u_1=u_2=1$, $u_{12}=0.95$ and $\chi=42.4$. }\label{Fig-Exp}
\end{figure}

For field higher than $B_{c2}$, the SC order is completely
suppressed. Among those terms in Eq.(\ref{F}), only the AF term
$\mathcal{F}_m$ and the magnetic field energy term survive. The
problem becomes uniform and can be solved analytically to yield
\begin{eqnarray}
  m(B)=\sqrt{|r_2|/u_2}\sqrt{1-B^2/B_s^2}  \label{canting-Eq}
\end{eqnarray}
where
\begin{eqnarray}
  B_s = \sqrt{|r_2|/\chi }~\phi_0/a^2 \label{Bs}
\end{eqnarray}
is the critical field where the AF moment disappears.

In Fig.\ref{Fig-Exp}, we plot the fit of the experimental results
of Matsuura{\it et al}\cite{matsuura2003,matsuura-unpublished}. Both the intrinsic AF
moment of NCCO and extrinsic FM moment of $Nd_2O_3$ impurity phase
contribute to the $(\pi,\pi,0)$ peak measured in $HK0$ zone
[corresponding to field in $(0,0,1)$ direction]\cite{MANG2003}
. However, the extrinsic contribution of the $Nd_2O_3$
impurity phase can be easily subtracted. Due to the cubic symmetry
of $Nd_2O_3$ crystal, we can subtract the integrated intensity of
$(0,0,4.4\pi)$ peak measured in $HHL$ zone, which arises
exclusively from $Nd_2O_3$, from the normalized integrated
intensity of $(\pi,\pi,0)$ measured in $HK0$ zone, to obtain the
intrinsic $NCCO$ contribution of the AF
moment\cite{matsuura-private}. We then take the square root of the
integrated intensity to obtain the moment. We obtain a reasonably
well fit of the field dependence of AF moment. For the field below
$B_{c2}$, our theory predicts that the FM moment per vortex scales
linearly with the field. It follows that the total FM moment
scales quadratically since the number of vortices is proportional
to the field. The experiment observed a linear relation. We
believe the quadratic relation between FM moment and field is
overwhelmed by the contribution of $Nd$ moment, which is linearly
polarized by the field.

The quantitative fit to the field dependence of the AF moment
determines $u_{12}/\sqrt{u_2u_1}=0.95$. Therefore, the quartic
terms are approximately $SO(5)$ symmetric. The quadratic terms
$r_1=-1$ and $r_2=-0.85$ deviate more from the $SO(5)$ symmetric
value. This is to be expected, since the system is close to
optimal doping where the SC state is the stable ground state.
Since $r_1$ depends on the chemical potential, it can always be
tune to $r_2$ at the quantum phase transition point where both
quadratic and quartic terms are approximately $SO(5)$ symmetric.
The experimental result of the field dependence of AF moment
beyond $B_{c2}$ can be fitted very well by the simple formula
Eq.(\ref{canting-Eq}) with a critical field $B_s\sim 20.5T$.
Energetically, this critical field is determined by the
competition between the Zeeman energy and the condensation energy
of the AF moment. From the fact that the observed AF moment at
$B_{c2}$ is about $m=0.05\mu_B$ per site\cite{Kang2003}, we
estimate that the Zeeman energy at $B_s$ is about $B_s m\approx
0.05 mev$. It follows that the AF coupling $J_{AF}$ is about
$20mev$, if the AF condensation energy can be written as
$E_{AF}=J_{AF} m^2$.

The detailed agreement between our model and the experiment
obtained above allows us to make a striking quantitative
prediction. As we mentioned previously, the upper critical field
$B_{c2}$ is strongly reduced by the repulsion between AF and SC.
Using the parameters obtained above, we estimate the bare value of
the SC coherence value to be $\xi_0^2=\rho_1/|r_1|$,
and the bare upper critical field is given by $B_0=\phi_0
|r_1|/2\pi\rho_1 \approx 23 T$. If the AF order could be
suppressed without disturbing the SC order, the SC order would be
enhanced. Consequently, the upper critical field $B_{c2}$ would
increase too. Our idea is to apply a Zeeman field in the
$ab$-plane. The Zeeman field suppress the AF moment through the
canting term $\chi$. After including the Zeeman field, the total
field is
\begin{eqnarray}
  {\bf b}=  b_z\hat{{\bf z}} + b_\parallel \hat{{\bf x}}.
\end{eqnarray}
The dimensionless upper critical field $b_{c2}$ is then given by
the solution of the coupled equations:
\begin{subequations}
\begin{eqnarray}
  \sqrt{b_{c2}^2- b_\parallel^2}=
   \frac{|r_1|-u_{12}m^2}{2\pi\rho_1/a^2} \label{bc2-1} 
\end{eqnarray}
\begin{eqnarray}
  m^2 = max\left[\frac{|r_2|-\chi b_{c2}^2}{u_2},0\right].\label{bc2-2}
\end{eqnarray}
\end{subequations}
The first equation gives the $z$-component of $b_{c2}$ as a
function of $m$ while the second one gives the AF moment $m$ at
$b_{c2}$.
\begin{figure}[t]
    \includegraphics[scale=0.9]{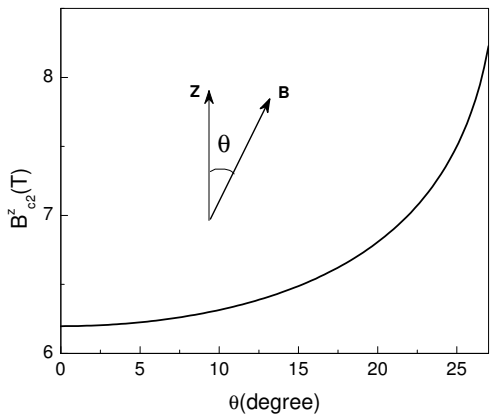}
    \caption{The angle dependence of the $z$-component of
    the upper critical field, given by the solution to coupled equations
    (\ref{bc2-1}) and (\ref{bc2-2}).
    The parameters are the same as those used in fitting of the
    experimental data in Fig.\ref{Fig-Exp}.}\label{Fig-Bc2}
\end{figure}
For small small $b_\parallel$, the increase of the $z$-component
of $b_{c2}$ is quadratic in $b_\parallel$
\begin{eqnarray}
 \frac{b_{c2}^z(b_\parallel)-b_{c2}^z(0)}{b_{c2}^z(0)}=
  \frac{\chi~ u_{12}b_{\parallel}^2+O(b_{\parallel}^3)}
  {u_2|r_1|-u_{12}|r_2|-\chi u_{12} (b_{c2}^z(0))^2  }
\end{eqnarray}
where $b_{c2}^z(0)$ is the upper critical field if
$b_\parallel=0$, given by the solution of self-consistent equation
\begin{eqnarray}
 b_{c2}^z(0)= \frac{u_2|r_1|-u_{12}|r_2|+\chi u_{12} (b_{c2}^z(0))^2}
  {2\pi u_2\rho_1/a^2}.
\end{eqnarray}

This effect can be used to quantitatively measure the competition
between the AF and the SC order parameters. Fig.\ref{Fig-Bc2}
plots the $z$-component of the upper critical field for NCCO
system studied in previous paragraphs, as a function of the
canting angle $\theta=\tan^{-1}(b^z_{c2}/b_\parallel)$. It is
interesting to note that a canting of $27^o$ increases the
$B^z_{c2}$ by about $30\%$. In the above estimate we neglected the
effect of the $ab$ plane field on the SC. If this effect is taken
into account, the actual increase of $B^z_{c2}$ would be smaller.

Similarly, for a system with {\it uniformly} co-existing AF order
and SC order at zero field, the suppression of the AF ordering due
to the Zeeman field in the $ab$ plane would increase the
transition temperature $T_c$ itself, according to this formula:
\begin{eqnarray}
\frac{T_c(b_\parallel)-T_c(0)}{T_c(0)}=
\frac{\chi u_{12}b_\parallel^2}{u_2|r_1|-u_{12}|r_2|}.
\end{eqnarray}
Here, we assume $r_2(T)=r_2(1-T/T_N^0)$, $r_1(T)=r_1(1-T/T_c^0)$ with $T_N^0>T_c^0$.
We also assume that the Zeeman field $b_\parallel$ is small enough
so that the AF moment $m$ is nonzero at $T_c$.
Again, the enhancement scales quadratically with the
Zeeman field.

In summary, we have studied the competition of AF and SC orders in
High-Tc superconductors by solving the GL free energy. We showed
that the curvature of the AF moment versus B field plot directly
determines the departure from the $SO(5)$ condition in the quartic
terms of the GL functional. Reasonable agreement with the recent
neutron scattering experiment on $Nd_{1.85}Ce_{0.15}CuO_4$ shows
that this system is close to a $SO(5)$ symmetric quantum critical
point. New experiments are predicted to increase $B_{c2}$ and
$T_c$ by applying an in-plane magnetic field. The GL model with
competing AF and SC order parameters can be derived from the
projected $SO(5)$ model on the lattice\cite{zhang1999}, which can
in turn be derived from the microscopic $t$-$J$ model by a
contractor-renormalization-group algorithm\cite{altman2002}. The
quantitative agreement between these approximately $SO(5)$
symmetric models and the experiment shows a promising direction
towards a full microscopic theory of high Tc superconductivity.

We would like to acknowledge useful discussions with Dr. M. R.
Beasley, B.A. Bernevig, Dr.P.C. Dai, Dr.E. Demler, Dr.J.P. Hu, Dr.H.J. Kang,
L. Lu, P. Mang, Dr.J. Zaanen, G. Zeltzer. This work was performed on behalf of
the US Department of Energy, Office of Basic Energy Sciences under
contract DE-AC03-76SF00515 and the National Science Foundation
under grant numbers DMR-9814289. HDC and CJW are also supported by
Stanford Graduate Fellowship.

%\bibliography{so5}

\end{document}